\documentclass[11pt]{article}

\usepackage[a4paper,margin=2.4cm]{geometry}
\usepackage[T1]{fontenc}
\usepackage[utf8]{inputenc}
\usepackage{lmodern}
\usepackage{microtype}
\usepackage{graphicx}
\usepackage{booktabs}
\usepackage{multirow}
\usepackage{siunitx}
\usepackage{amsmath}
\usepackage{amssymb}
\usepackage{xcolor}
\usepackage{hyperref}
\usepackage{cleveref}
\usepackage[backend=biber,style=numeric-comp,sorting=none,giveninits=true,maxbibnames=99]{biblatex}
\usepackage{enumitem}
\usepackage{lineno}

\hypersetup{colorlinks=true,citecolor=blue,linkcolor=blue,urlcolor=blue}
\newcommand{\Dplus}{\mathrm{D}^{+}}
\newcommand{\nsep}{n_{\Dplus,\mathrm{sep}}}
\newcommand{\Pin}{P_{\mathrm{in}}}
\newcommand{\GN}{\Gamma_{\mathrm{N}}}
\newcommand{\Te}{T_{\mathrm{e}}}

\title{Machine learning surrogates for multi-geometry EDGE2D--EIRENE studies of MAST Upgrade}
\author{%
\parbox{0.9\textwidth}{%
\centering
George K. Holt\textsuperscript{1,$\dagger$},
Lorenzo Zanisi\textsuperscript{1},
Romain Futtersack\textsuperscript{1},
James Simpson\textsuperscript{1}
\\[0.5em]
\small \textsuperscript{1}UKAEA (United Kingdom Atomic Energy Authority), Culham Campus, Abingdon, Oxfordshire OX14 3DB, United Kingdom
\\[0.5em]
\small \textsuperscript{$\dagger$}\texttt{george.holt@ukaea.uk}
}}
\date{}

\begin{document}
\maketitle

\begin{abstract}
Two-dimensional edge plasma transport simulations provide a detailed description of divertor plasma behaviour, but their computational cost limits their routine use in large parameter scans and iterative modelling workflows.
We present a database of \num{23379} converged EDGE2D--EIRENE simulations for MAST Upgrade (MAST-U) spanning conventional divertor (CD), elongated divertor (ED) and Super-X divertor (SXD) magnetic configurations, experimentally relevant ranges of separatrix density and input power, multiple deuterium fuelling locations, and nitrogen seeding of varying amount and puff location.
Independent multilayer perceptron surrogate models are trained for twelve representative scalar outputs, including divertor front positions, target electron temperature, core-edge boundary quantities, the deuterium fuelling rate required to maintain a prescribed separatrix density, and total deuterium ion content.
We demonstrate two applications: comparing the density dependence of carbon radiation front motion across divertor geometries and fuelling locations, and assessing the response of divertor fronts, target temperature and core-boundary temperature to nitrogen seeding rate and location.
On a held-out test set drawn uniformly and at random from the domain, the surrogate models reproduce EDGE2D--EIRENE with coefficients of determination of $0.9462\leq R^2\leq0.9994$, including root-mean-square errors (RMSEs) of 0.0303 in normalised carbon radiation front position and \SI{0.108}{eV} in peak lower outer divertor target electron temperature.
An ordinary least squares linear regression baseline with the same inputs gives $0.0766\leq R^2\leq0.9879$ and 3.0--12.2 times larger RMSEs.
Dense surrogate scans at fixed input power quantify how the upstream motion of the carbon radiation front with increasing separatrix density depends on divertor geometry and deuterium fuelling location, including differences in the density at which rapid front displacement occurs.
Similarly, nitrogen seeding scans at fixed density and power reveal a pronounced dependence on seeding location: high-field-side seeding at rates of approximately \(3\times10^{20}\)–\(6\times10^{20}\,\mathrm{s^{-1}}\) rapidly reduces the core boundary electron temperature in all three configurations, whereas divertor seeding produces little response.
Single-point inference for one scalar requires \SI{1.86}{\milli\second} on one CPU thread and \SI{1.82}{\milli\second} on GPU, while batch inference for all twelve outputs at \num{786432} operating points takes \SI{1.76}{\second} on GPU.
A separate set of EDGE2D--EIRENE runs had a median wall time of \SI{2.61}{\hour} using four CPU cores.
The models therefore enable rapid comparison of geometry and source-location effects on divertor and core-boundary conditions within the sampled EDGE2D--EIRENE operating domain.
\end{abstract}

\section{Introduction}
\label{sec:introduction}

Power exhaust is a central challenge for magnetically confined fusion devices~\cite{asakura2023}.
Heat and particles leaving the confined plasma are transported through the scrape-off layer (SOL) to plasma-facing components, where excessive steady state and transient loads can constrain both plasma performance and component lifetime~\cite{loarte2017,Krieger2025}.
Divertor detachment, in which volumetric power and momentum losses reduce the plasma temperature and heat flux at the target, is therefore an important element of reactor-relevant exhaust studies~\cite{krasheninnikov2017}.
The onset of detachment, as well as location and stability of the detachment front, depend on a coupled set of geometric and operational parameters, including divertor magnetic configuration, upstream density, input power, fuelling and impurity radiation~\cite{lipschultz2016,lee2026,xia2026}.
Conditions in the SOL can also have a strong effect on the confined plasma.
For example, strongly radiating SOL regimes can be deleterious for confinement~\cite{wang2017}, while X-point radiator configurations have been observed to enable operation in highly dissipative regimes~\cite{bernert2025}.
Furthermore, fuel introduced by gas puff must transport through the SOL in order to breach the core, while seeded impurities are often introduced to modulate SOL conditions but may themselves transport to the core and dilute fusion performance.

MAST Upgrade (MAST-U) was designed with broad divertor shaping capability, permitting conventional and long-legged configurations to be compared while retaining similar core plasma shapes~\cite{fishpool2013,morris2018}.
Recent experiments have demonstrated strong effects of divertor shaping and neutral baffling on target conditions and detachment access~\cite{wijkamp2023,verhaegh2025}.
The flexibility of MAST-U therefore makes it a useful test bed for understanding how changes to divertor geometry interact with operational controls such as fuelling and impurity seeding~\cite{verhaegh2023,verhaegh2024,kool2025}.
Earlier SOLPS modelling similarly predicted substantial differences between conventional and Super-X configurations through changes in flux expansion, connection length, volumetric losses and neutral behaviour~\cite{havlickova2014}.
Practical questions include how the density required for a given divertor front displacement changes with configuration and fuelling location, and how impurity seeding affects target cooling alongside the plasma state at the core boundary.

Interpretation and prediction of these effects commonly rely on two-dimensional edge plasma transport codes.
For example, earlier SOLPS modelling predicted substantial differences between conventional and Super-X configurations through changes in flux expansion, connection length, volumetric losses and neutral behaviour~\cite{havlickova2014}.
A similar code is EDGE2D--EIRENE, which couples the multi-fluid EDGE2D plasma transport model to the EIRENE kinetic Monte Carlo neutral transport code~\cite{simonini1994,reiter2005,moulton2011}.

Recycled and injected neutrals can travel away from their source before ionisation, charge exchange or molecular reactions, so their transport helps determine where particles, momentum and energy are exchanged with the plasma~\cite{reiter2005}.
Experiments and modelling show that divertor baffling and fuelling location affect neutral retention and detachment access~\cite{verhaegh2025,lee2026}.
This spatially resolved coupling is pertinent to many quantities of interest in tokamak experiments and design, including target temperatures, divertor front positions and deuterium fuelling requirements.
Resolution of neutral transport effects is therefore critical for understanding the interplay between divertor geometry, fuelling and impurity seeding.

The cost is that a single calculation can require many hours to converge to steady state, with numerical convergence becoming particularly difficult in strongly radiating conditions and in configurations that exhibit steep profile gradients.
Direct simulation is consequently poorly suited to dense scans, optimisation, uncertainty quantification or other workflows requiring large numbers of model evaluations.

Surrogate modelling provides a complementary approach.
A surrogate is trained on a finite database of high-fidelity simulations and subsequently interpolates between those calculations at negligible cost relative to the parent model.
Previous work has demonstrated this concept for divertor and SOL modelling using one-dimensional UEDGE~\cite{zhu2022}, Hermes-3 calculations~\cite{holt2024} and DIV1D~\cite{poels2023}, reduced fidelity SOLPS--ITER databases~\cite{dasbach2023}, and two-dimensional UEDGE~\cite{zhu2025,zhao2026} and SOLPS~\cite{gopakumar2020} databases.
These studies establish that edge plasma quantities can be emulated accurately enough for rapid analysis and, in some cases, applications oriented towards divertor control.
However, end-to-end surrogate studies published to date have predominantly used either one-dimensional plasma--neutral models or two-dimensional databases with reduced neutral and/or impurity treatments.
In particular, a surrogate based on a two-dimensional plasma--neutral model with kinetic neutrals and self-consistent impurity treatment is lacking for MAST-U, despite the machine's focus on divertor studies where neutral effects play a crucial role.

The present work addresses this gap using a database of MAST-U EDGE2D--EIRENE simulations covering three magnetic configurations and a broad operational parameter space.
Simple multilayer perceptrons (MLPs) are trained independently for twelve scalar outputs.
Two applications demonstrate how these models can be used to investigate SOL physics.
First, unseeded density scans at fixed input power compare carbon radiation front motion across the three configurations and four deuterium fuelling locations, quantifying differences in the density required to reach a given normalised front position (section~\ref{sec:scan}).
Second, nitrogen seeding scans at fixed density and power compare the response to seeding rate and puff location using carbon radiation and electron temperature fronts, target electron temperature and core-boundary electron temperature (section~\ref{sec:nitrogen-scan}).
Comparisons with nearby converged EDGE2D--EIRENE cases assess the scan trends, with held-out cases providing independent checks of surrogate accuracy.
These applications use interpolation within the sampled simulation domain under fixed transport assumptions.
Broader applications are discussed in section~\ref{sec:future}.

\section{EDGE2D--EIRENE simulation database}
\label{sec:database}

\subsection{MAST-U configurations and EDGE2D--EIRENE model}
\label{sec:model}

EDGE2D solves a two-dimensional multi-fluid description of the edge plasma on a flux-aligned computational mesh~\cite{simonini1994}.
Neutral transport is treated by EIRENE using a kinetic Monte Carlo approach, with the plasma and neutral solutions coupled iteratively~\cite{reiter2005,moulton2011}.
In the simulations considered here, deuterium is the main plasma species, nitrogen is optionally included as an externally seeded impurity, and sputtered carbon is treated self-consistently within the EDGE2D--EIRENE setup and always included.

\begin{figure}[htbp]
    \centering
    \includegraphics[width=\linewidth]{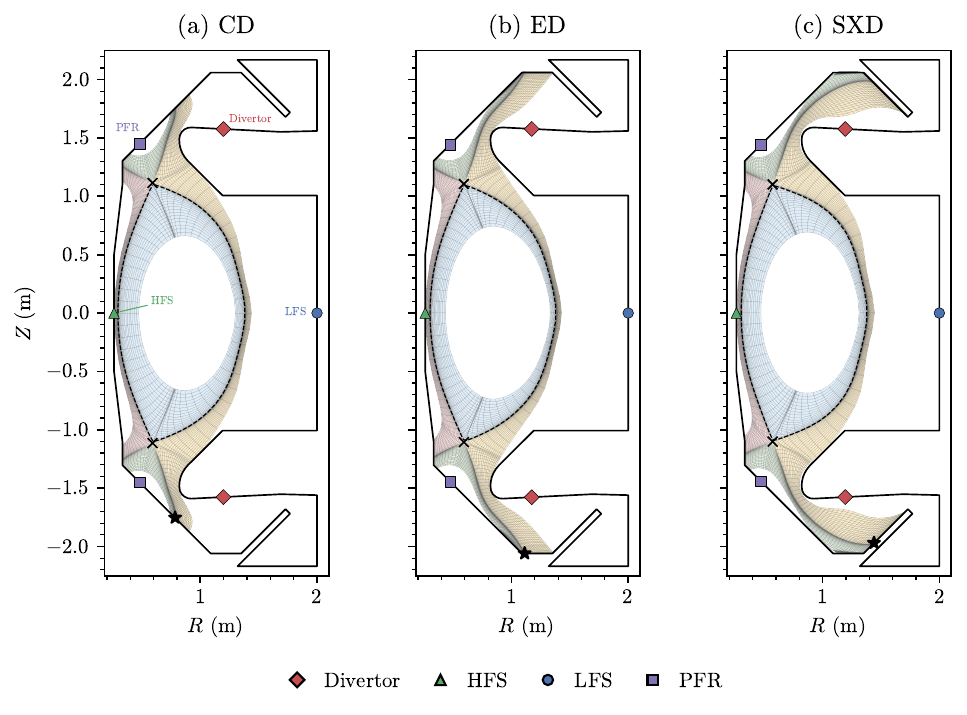}
    \caption{MAST-U vessel and divertor outlines used in the simulations (black) and the EDGE2D--EIRENE computational meshes for (a) conventional divertor (CD), (b) elongated divertor (ED), and (c) Super-X divertor (SXD).
    Grid cells are shaded by computational domain: core (blue), inner SOL (red), outer SOL (yellow), and private flux region (green).
    Dashed curves denote separatrices, crosses X-points, and stars the lower outer strike points.
    Markers show representative centres of the four deuterium-fuelling and impurity seeding source regions.}
    \label{fig:grids}
\end{figure}

Three magnetic configurations are represented: a conventional divertor (CD) based on MAST-U discharge 45470, an elongated divertor (ED) based on discharge 47079, and a Super-X divertor (SXD) based on discharge 46860.
These equilibria have broadly similar core shapes but markedly different outer divertor leg geometries.
\Cref{fig:grids} shows the three computational meshes and the locations used to define localised gas fuelling and impurity seeding.
The grids used for CD simulations are the same as that used in~\textcite{moulton2024}, while the ED and SXD grids are those used in~\textcite{verhaegh2025}.

The anomalous cross-field transport coefficients are held fixed across the complete database.
This is an important restriction on the interpretation of the sampled domain: the database explores geometry and operational variation conditional on the selected transport model.
All modelled ion species used a spatially uniform particle diffusivity of $D_\perp=\SI{4}{\metre\squared\per\second}$.
The electron and ion thermal diffusivities were identical: along the outer mid-plane, $\chi_\mathrm{e}=\chi_\mathrm{i}=\SI{10}{\metre\squared\per\second}$ for $r\leq\SI{-1}{\milli\metre}$, decreased linearly to \SI{4}{\metre\squared\per\second} at the separatrix ($r=0$), and remained \SI{4}{\metre\squared\per\second} in the SOL ($r\geq0$).
Both thermal diffusivities were also \SI{4}{\metre\squared\per\second} in the private-flux region.
Here $x$ is the signed radial distance from the separatrix along the outer mid-plane, with positive values in the SOL.
The core/SOL profile was applied as a flux-surface-constant coefficient.

Neutral particles reaching the EIRENE core boundary were absorbed, with the absorbed particle flux returned as an ion influx through the EDGE2D core boundary.
The hydrogenic albedo at the main chamber and private region walls was 0.999.
Deuterium recycled at the targets with unit particle albedo.
Recycled neutrals were thermally re-emitted, launched at an angle sampled from a cosine angular distribution about the surface normal.
Two additional divertor pump surfaces had albedo 0.9712 for atoms, molecules and ions.
EIRENE impurity recycling was enabled for carbon and disabled for nitrogen, and neutral--neutral collisions were disabled.
Atomic and molecular coefficients were taken from EIRENE Database v2.0 and the ADAS v1.3 ADF11 collection.

Earlier EDGE2D--EIRENE modelling of JET matched measured outer midplane profiles and outer target heat load distribution, with calculated parallel saturation current within Langmuir probe tolerance~\cite{moulton2011}.
Related SOLPS--ITER modelling of MAST-U conventional and Super-X divertor discharges reproduced the observed weak effect of divertor configuration on upstream profiles and the movement of detachment fronts~\cite{moulton2024}.
The present surrogate accuracy metrics quantify agreement with the EDGE2D--EIRENE simulations used for training. Experimental accuracy of these particular setups is not assessed here, but we note that the physics models underpinning EDGE2D--EIRENE and SOLPS--ITER have been well validated for MAST-U.

\subsection{Parameter space}
\label{sec:parameter-space}

The database varies six inputs: the outer-midplane separatrix deuterium ion density $\nsep$, magnetic geometry, total input power $\Pin$, deuterium fuelling location, nitrogen puff rate $\GN$, and nitrogen puff location.
Their ranges are summarised in \cref{tab:inputs}.
The continuous ranges span the current MAST-U operational regime.
The geometry and gas source locations are discrete categories chosen to represent physically distinct configurations available in MAST-U.
Input power is divided equally between electron and ion energy channels.
Deuterium and nitrogen sources can each be placed at the low-field side (LFS), high-field side (HFS), private flux region (PFR) or divertor.
Fuelling and seeding (where present) locations are up-down symmetric.

The deuterium puff rate is intentionally not prescribed as an input.
Instead, a proportional--integral--derivative (PID) controller adjusts the deuterium source during each EDGE2D--EIRENE calculation to achieve the target outer-midplane $\nsep$.
The converged deuterium puff rate is consequently a simulation output and is one of the quantities emulated in section~\ref{sec:surrogate-outputs}.

\begin{table}[t]
\centering
\caption{Input parameter space used to generate the EDGE2D--EIRENE database and its representation in the MLP surrogate. Continuous inputs are standardised using statistics calculated from the training data only.}
\label{tab:inputs}
\begin{tabular}{llll}
\toprule
Input & Symbol & Type & Range/categories \\
\midrule
Outer midplane density & $\nsep$ & Continuous & $\SI{5e18}{m^{-3}}$--$\SI{2e19}{m^{-3}}$ \\
Magnetic geometry & -- & Categorical & CD, ED, SXD \\
Core boundary power & $\Pin$ & Continuous & $\SI{0.5}{MW}$--$\SI{3.0}{MW}$ \\
D fuelling location & -- & Categorical & LFS, HFS, PFR, divertor \\
N puff rate & $\GN$ & Continuous & $0$--$\SI{2e21}{s^{-1}}$ \\
N puff location & -- & Categorical & LFS, HFS, PFR, divertor \\
\bottomrule
\end{tabular}
\end{table}

\subsection{Sampling and simulation workflow}
\label{sec:sampling}

A total of \num{38412} EDGE2D--EIRENE calculations were launched.
An initial coarse sampling of the parameter space was generated using a Sobol low-discrepancy sequence~\cite{sobol1967}.
This first tranche was intended to provide broad coverage and, critically, a bank of converged states from which subsequent calculations could be initialised.
The parameter space was then populated more densely using random sampling.
For each new non-Sobol calculation, the nearest converged Sobol-sampled case in input space was selected as the initial state.
This strategy was used to reduce the transient required to approach the requested solution and to improve the probability of convergence within the available computational budget.

\begin{figure}[t]
    \centering
    \includegraphics[width=\linewidth]{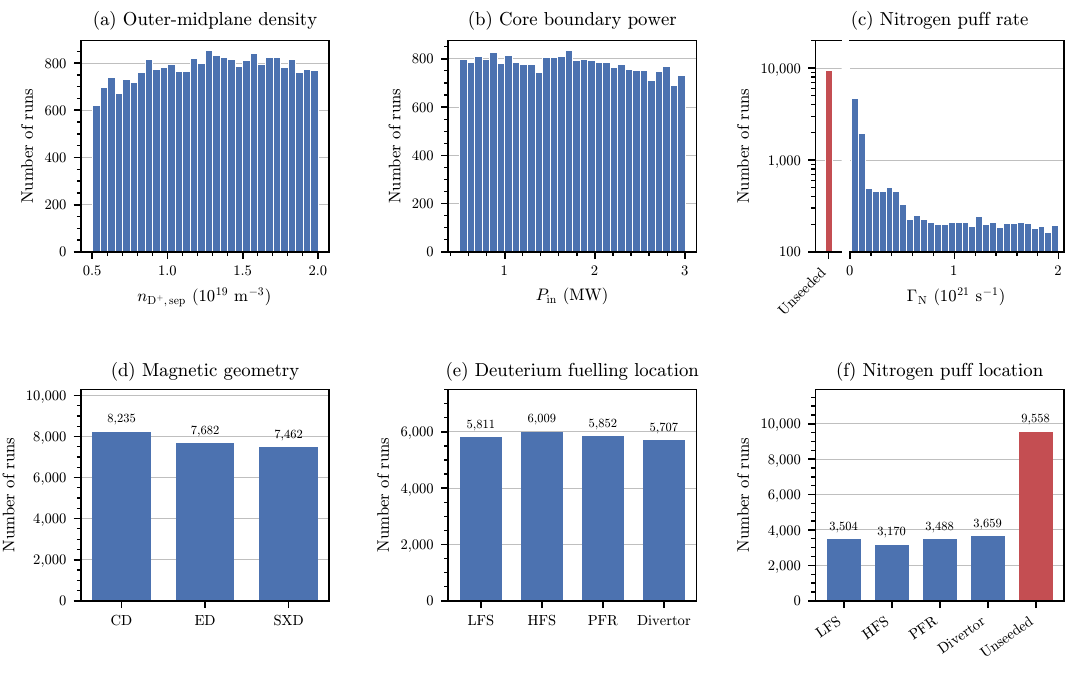}
    \caption{Distributions of the six input parameters across the \num{23379} converged EDGE2D--EIRENE simulations: (a) outer-midplane separatrix deuterium ion density $\nsep$, (b) core boundary input power $\Pin$, (c) nitrogen puff rate $\GN$, (d) magnetic geometry, (e) deuterium fuelling location and (f) nitrogen puff location.
    The red bars in (c) and (f) represent the \num{9558} unseeded cases ($\GN=0$).}
    \label{fig:converged_input_distributions}
\end{figure}

The simulation campaign was generated progressively.
Cases with deuterium as the only introduced species were run first, followed by nitrogen seeded cases in tranches of increasing seeding rate.
Approximately 40.9\% of the complete simulation set is deuterium-only, with carbon included self-consistently.
Calculations are distributed approximately uniformly across the three magnetic geometries and across the available fuelling and seeding locations.
Each simulation used four CPU cores and was allocated a maximum wall time of \SI{48}{\hour}.
Runs that did not satisfy the convergence criteria (described in~\Cref{sec:convergence}) within this budget were excluded from surrogate training.
Of the \num{38412} launched cases, \num{23379} were retained as converged.
The convergence rate is high for deuterium-only cases and decreases as nitrogen seeding is increased, reflecting the greater numerical difficulty of strongly radiating conditions.
It is noted that the vast majority of unconverged cases with nitrogen seeding failed to converge during the allocated computational budget, and that they did not fail numerically.
It is therefore likely that increased computational resources would allow for more converged nitrogen cases to be added to the database in the future.
The parameter space coverage of the converged cases is shown in \cref{fig:converged_input_distributions}.

\begin{figure}[t]
    \centering
    \includegraphics[width=\linewidth]{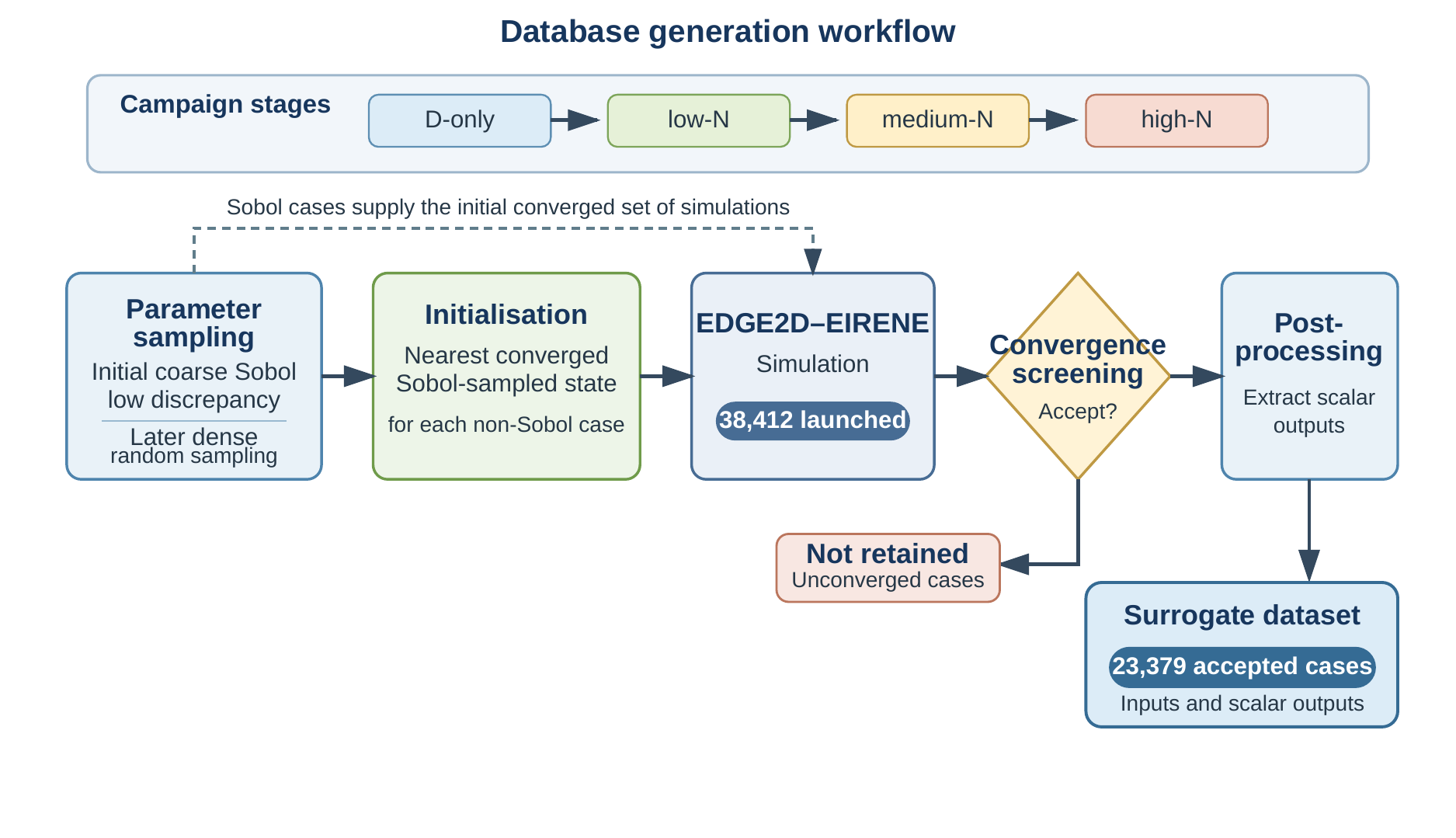}
    \caption{Database generation workflow.
    An initial Sobol sample supplied a set of converged initial states for the subsequent randomly sampled cases, which were initialised from the nearest converged Sobol case in input space.
    The campaign progressed from deuterium-only to increasingly seeded nitrogen cases.
    Of the \num{38412} launched EDGE2D--EIRENE calculations, \num{23379} passed convergence screening and were post-processed to form the surrogate dataset.}
    \label{fig:workflow}
\end{figure}

\Cref{fig:workflow} summarises the database generation process.
The combination of initial low-discrepancy sampling, reuse of converged states and automated convergence screening enabled the simulation campaign to be conducted with limited manual intervention.
This automated workflow is also important for future expansion of the database to additional transport assumptions, equilibria and other input quantities.

\subsection{Convergence screening}
\label{sec:convergence}

Convergence was assessed automatically from three slowly evolving quantities: the flux surface averaged ion temperature at the core boundary, the flux surface averaged electron temperature at the core boundary, and the total deuterium ion content in the EDGE2D computational domain.
For each signal, a smoothed and normalised temporal gradient is converted to a post-convergence probability using a binary logistic regression classifier trained on manually annotated convergence times from an independent representative dataset.
A simulation is accepted only when all three signals are classified as post-convergence.
The detailed screening algorithm is given in appendix~\ref{app:convergence}.

\subsection{Surrogate output quantities}
\label{sec:surrogate-outputs}

The converged EDGE2D--EIRENE database contains sufficient information to construct surrogate models for many scalar quantities without repeating or extending the underlying simulations.
The present work selects twelve outputs spanning divertor front locations, target conditions, core-boundary plasma state, global content and the fuelling requirement imposed by density control.
These quantities are listed in \cref{tab:outputs}.
They are intended to demonstrate the breadth of quantities that can be emulated rather than to define an exhaustive surrogate output list.
The cost of adding new surrogate outputs is small.

Front positions are expressed in terms of normalised poloidal distance along the lower outer separatrix, with $s_{\mathrm{norm}}=0$ at the target and $s_{\mathrm{norm}}=1$ at the X-point.
The front search includes mesh points from the target up to, but excluding, the X-point sample.
The carbon radiation front is located at the mesh point where the magnitude of carbon radiation is greatest.
Separately, temperature fronts are defined at a threshold $T_*\in\{1,3\}\,\mathrm{eV}$. The temperature front is located at the mesh point that minimises $|\Te-T_*|$, without interpolation between points.
The same definition applies when $T_*$ lies outside the sampled temperature range. The reported position then corresponds to the closest available temperature.

Core boundary quantities are sampled in the core-facing cells of the first SOL rings on the LFS and HFS, between the two X-points and excluding the divertor legs.
For deuterium and impurity ion densities and electron and ion temperatures, the volume-weighted mean is defined as $\langle q\rangle_{\mathrm{cb}}=\sum_j q_j\Delta V_j/\sum_j\Delta V_j$, where $j$ indexes cells in these portions of the rings and $\Delta V_j$ is the cell volume.
The impurity densities include the sum over all charge states of each element before this spatial average is taken.

\begin{table}[htbp]
\centering
\caption{Scalar EDGE2D--EIRENE quantities emulated in this work.}
\label{tab:outputs}
\begin{tabular}{@{}p{0.60\linewidth}p{0.18\linewidth}p{0.12\linewidth}@{}}
\toprule
Quantity & Symbol & Units \\
\midrule
Carbon radiation front: normalised position of maximum carbon radiation magnitude & $s_{\mathrm{C,rad}}$ & -- \\
Final PID-controlled deuterium neutral puff rate & $\Gamma_{\mathrm{D}}$ & \si{s^{-1}} \\
Volume-weighted mean carbon ion density at the core boundary & $\langle n_{\mathrm{C}}\rangle_{\mathrm{cb}}$ & \si{m^{-3}} \\
Volume-weighted mean nitrogen ion density at the core boundary & $\langle n_{\mathrm{N}}\rangle_{\mathrm{cb}}$ & \si{m^{-3}} \\
Electron density at the LFS separatrix outer midplane & $n_{\mathrm{e,LFS}}$ & \si{m^{-3}} \\
Volume-weighted mean deuterium ion density at the core boundary & $\langle n_{\Dplus}\rangle_{\mathrm{cb}}$ & \si{m^{-3}} \\
Peak electron temperature along the lower LFS divertor target & $\Te^{\mathrm{tgt,max}}$ & eV \\
Volume-weighted mean electron temperature at the core boundary & $\langle \Te\rangle_{\mathrm{cb}}$ & eV \\
Normalised position of the $\Te=\SI{1}{eV}$ front (nearest sampled $\Te$) & $s_{1\,\mathrm{eV}}$ & -- \\
Normalised position of the $\Te=\SI{3}{eV}$ front (nearest sampled $\Te$) & $s_{3\,\mathrm{eV}}$ & -- \\
Volume-weighted mean deuterium ion temperature at the core boundary & $\langle T_{\mathrm{i},\Dplus}\rangle_{\mathrm{cb}}$ & eV \\
Total deuterium ion content: sum of density times cell volume over the plasma domain & $N_{\Dplus}$ & particles \\
\bottomrule
\end{tabular}
\end{table}

\section{Neural network surrogate model}
\label{sec:surrogate}

\subsection{Emulation task and data split}
\label{sec:data-split}

A separate single-output MLP regressor is trained for each quantity in \cref{tab:outputs}.
This choice keeps the training problem simple and allows additional scalar quantities to be incorporated without changing the surrogate interface.
The six inputs listed in \cref{tab:inputs} are used for every model.
The \num{23379} converged simulations are split randomly into an outer training pool containing 80\% of the data and a held-out test set containing the remaining 20\%.
For MLP optimisation, 20\% of the outer training pool is used as a validation set and the remaining 80\% is used for gradient-based optimisation.

Geometry and gas source locations are one-hot encoded\footnote{Variables are represented as a binary vector in which each category corresponds to a unique position, with a value of 1 for the selected category and 0 for all others.}, and input features and regression targets are standardised to zero mean and unit variance.
Preprocessing parameters are estimated from the complete outer training pool before the split into optimisation and validation sets.
The held-out test set is excluded from this process.

\subsection{Network architecture and optimisation}
\label{sec:architecture}

All twelve surrogate models use the same architecture and training procedure.
Each model is a fully connected MLP with three hidden layers of 256 neurons and rectified linear unit (ReLU) activations.
The resulting single-output regressor contains \num{135937} trainable parameters.
No output-specific architecture search or hyperparameter optimisation is performed. The purpose is to test whether a single straightforward model class is sufficient across a heterogeneous set of plasma observables.

The networks are trained using AdamW~\cite{loshchilov2019} with a mean-reduced mean-squared-error (MSE) loss in standardised target space.
The initial learning rate is \num{1e-3}, the mini-batch size is 256, and the AdamW weight decay is \num{1e-4}.
The learning rate is reduced by a factor of 0.5 if the validation MSE does not improve for 20 epochs.
For early stopping, an improvement is defined as a reduction in validation MSE of more than \num{1e-5} relative to the previously accepted minimum.
Training stops after \num{50} epochs without such an improvement, and the parameters corresponding to the last minimum are retained.
Training is limited to \num{1000} epochs, although all models terminate earlier through early stopping.

\subsection{Linear regression baseline}
\label{sec:linear-baseline}

To assess whether a nonlinear model is required, an ordinary least-squares (OLS) linear regression with an intercept is fitted separately for each quantity in \cref{tab:outputs}.
The baseline uses the same six inputs, the same one-hot encoding of the categorical inputs and the same input standardisation as the MLPs.
It has no hyperparameters and therefore requires no validation set.
It is fitted to the complete outer training pool and evaluated on the same held-out test set as the MLPs.
Each output is regressed in physical units.
Standardising the target, as is done for the MLPs, would not change the OLS predictions.
Because the one-hot indicators for each categorical input sum to unity, the design matrix is rank deficient. That is to say, because the columns contain redundant information, there is not a unique set of regression coefficients.
We therefore use the minimum-norm least-squares solution, which selects the solution with smallest overall coefficient magnitudes from the choice of all equally-good solutions.

\subsection{Accuracy metrics}
\label{sec:metrics}

Surrogate accuracy is evaluated on the held-out test set in physical units.
For target values $y_i$, predictions $\hat{y}_i$ and $N$ test samples, the reported metrics are the root-mean-square error
\begin{equation}
    \mathrm{RMSE}=\sqrt{\frac{1}{N}\sum_{i=1}^{N}(\hat{y}_i-y_i)^2},
\end{equation}
the mean absolute error
\begin{equation}
    \mathrm{MAE}=\frac{1}{N}\sum_{i=1}^{N}|\hat{y}_i-y_i|,
\end{equation}
and the coefficient of determination
\begin{equation}
    R^2 = 1-\frac{\sum_i (y_i-\hat{y}_i)^2}{\sum_i (y_i-\bar{y})^2}.
\end{equation}
The outputs span different physical units and scales, so their absolute errors are not directly comparable.

\section{Surrogate accuracy}
\label{sec:accuracy}

\subsection{Accuracy across all outputs}
\label{sec:all-outputs}

\Cref{tab:performance} summarises held-out test performance for all twelve outputs.
The coefficients of determination range from \num{0.9462} for the \SI{1}{eV} temperature front position to \num{0.9994} for the LFS separatrix midplane electron density.
The carbon radiation and \SI{3}{eV} temperature front positions have RMSEs of \num{0.0303} and \num{0.0153}, respectively, in normalised poloidal distance, while the peak lower LFS target electron temperature has an RMSE of \SI{0.108}{eV}.
For the core boundary electron and deuterium ion temperatures, the RMSEs are \SI{0.815}{eV} and \SI{1.24}{eV}, respectively.
The \SI{1}{eV} temperature front position and total deuterium ion content are the most difficult of the reported outputs ($R^2=0.9462$ and \num{0.9664}, respectively).
The core boundary nitrogen density has the next lowest coefficient of determination ($R^2=0.9834$).

The OLS baseline defined in \cref{sec:linear-baseline} is less accurate than the MLP for every output.
The test RMSE of the baseline is larger by a factor of between \num{3.0}, for the total deuterium ion content, and \num{12.2}, for the peak lower LFS target electron temperature.
The baseline coefficients of determination range from \num{0.0766} for the core boundary carbon density to \num{0.9879} for the LFS separatrix midplane electron density, and are below \num{0.9} for ten of the twelve outputs.
The largest baseline coefficient of determination is obtained for $n_{\mathrm{e,LFS}}$, which is closely related to the prescribed outer midplane separatrix density $\nsep$, with a Pearson correlation of \num{0.985} between the two in the outer training pool.
The other output with baseline $R^2>0.9$ is the core boundary electron temperature ($R^2=0.9030$).
By contrast, the linear model captures less than half of the test set variance in the core boundary carbon and nitrogen densities, peak target electron temperature and \SI{1}{eV} front position ($R^2=0.0766$, $0.4645$, $0.2943$ and $0.4416$, respectively).
For every output, the training pool and test set coefficients of determination of the baseline differ by less than \num{0.014}, indicating avoiding of overfitting.
The larger errors therefore arise from the restricted functional form.
A linear model contains no interactions between the continuous inputs and the magnetic geometry or gas puff locations, and over the sampled domain most outputs cannot be represented accurately in this form.

\begin{table}[htbp]
\centering
\caption{Held-out test set performance of the MLP surrogate models, with the RMSE of the OLS linear regression baseline for comparison.
The rightmost column gives the minimum and maximum reference EDGE2D--EIRENE output values across the same \num{4676} test cases.
Output symbols and definitions are given in \cref{tab:outputs}; errors and reference values are reported in the listed units.}
\label{tab:performance}
\small
\setlength{\tabcolsep}{3pt}
\begin{tabular}{llccccc}
\toprule
 & & \multicolumn{3}{c}{MLP} & OLS & Test-set range \\
\cmidrule(lr){3-5}\cmidrule(lr){6-6}
Output & Units & MAE & $R^2$ & RMSE & RMSE & Min--max \\
\midrule
$s_{\mathrm{C,rad}}$ & -- & 0.0124 & 0.9892 & 0.0303 & 0.184 & $0.00394\text{--}0.981$ \\
$\Gamma_{\mathrm{D}}$ & \si{s^{-1}} & \num{3.76e20} & 0.9934 & \num{6.39e20} & \num{3.95e21} & $(0\text{--}3.34)\times10^{22}$ \\
$\langle n_{\mathrm{C}}\rangle_{\mathrm{cb}}$ & \si{m^{-3}} & \num{1.78e15} & 0.9887 & \num{3.54e15} & \num{3.20e16} & $(1.32\text{--}22.0)\times10^{16}$ \\
$\langle n_{\mathrm{N}}\rangle_{\mathrm{cb}}$ & \si{m^{-3}} & \num{4.85e15} & 0.9834 & \num{1.74e16} & \num{9.87e16} & $(0\text{--}1.66)\times10^{18}$ \\
$n_{\mathrm{e,LFS}}$ & \si{m^{-3}} & \num{4.12e16} & 0.9994 & \num{9.93e16} & \num{4.52e17} & $(0.518\text{--}2.26)\times10^{19}$ \\
$\langle n_{\Dplus}\rangle_{\mathrm{cb}}$ & \si{m^{-3}} & \num{4.45e17} & 0.9968 & \num{1.14e18} & \num{9.93e18} & $(0.554\text{--}10.2)\times10^{19}$ \\
$\Te^{\mathrm{tgt,max}}$ & eV & 0.0280 & 0.9952 & 0.108 & 1.32 & $0.585\text{--}19.5$ \\
$\langle \Te\rangle_{\mathrm{cb}}$ & eV & 0.294 & 0.9979 & 0.815 & 5.48 & $1.14\text{--}74.5$ \\
$s_{1\,\mathrm{eV}}$ & -- & 0.0170 & 0.9462 & 0.0587 & 0.189 & $0.00213\text{--}0.981$ \\
$s_{3\,\mathrm{eV}}$ & -- & 0.00801 & 0.9976 & 0.0153 & 0.139 & $0.00394\text{--}0.981$ \\
$\langle T_{\mathrm{i},\Dplus}\rangle_{\mathrm{cb}}$ & eV & 0.489 & 0.9986 & 1.24 & 11.8 & $1.15\text{--}158$ \\
$N_{\Dplus}$ & particles & \num{5.99e17} & 0.9664 & \num{2.65e18} & \num{7.96e18} & $(0.729\text{--}9.13)\times10^{19}$ \\
\bottomrule
\end{tabular}
\end{table}

\subsection{Representative prediction quality}
\label{sec:representative-accuracy}

\begin{figure}[htbp]
    \centering
    \includegraphics[width=\linewidth]{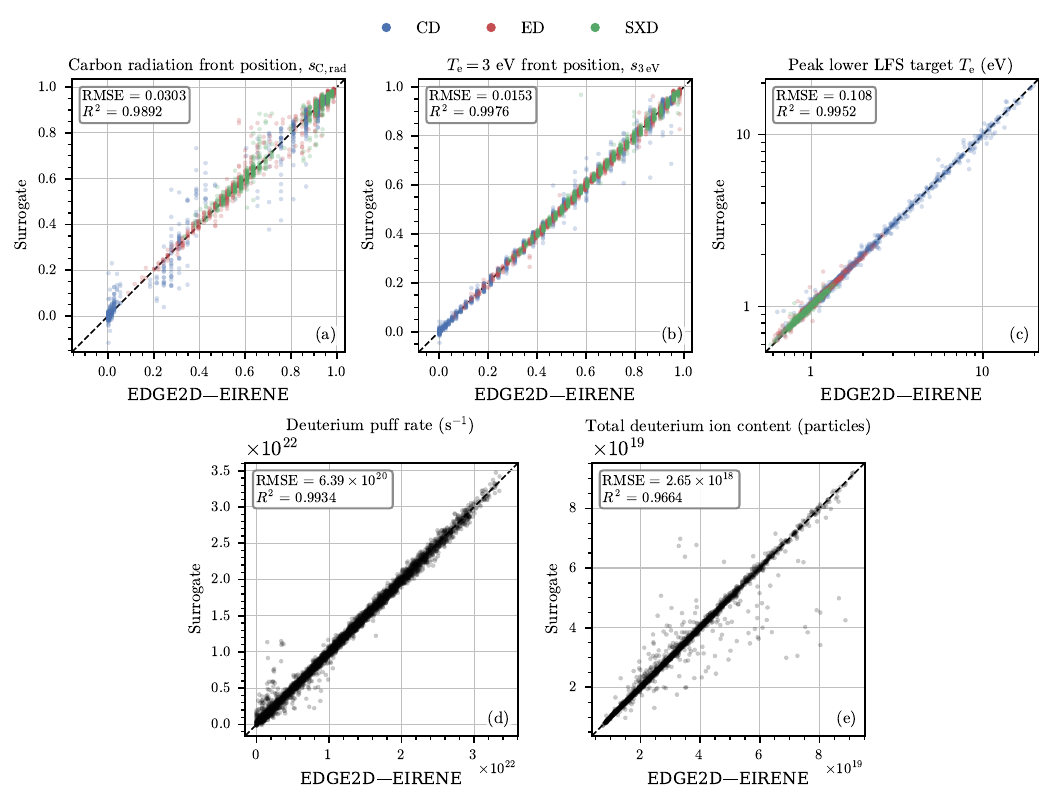}
    \caption{Parity plots for held-out test samples ($n=4676$) for (a) carbon radiation front position, (b) $\Te=\SI{3}{eV}$ front position, (c) peak lower LFS target $\Te$, (d) deuterium puff rate and (e) total deuterium ion content.
    Colours distinguish conventional (CD), elongated (ED) and Super-X (SXD) divertor configurations in panels (a)--(c), whereas black points are used in panels (d)--(e) because their geometry distributions overlap substantially.
    The dashed line denotes equality, and each panel reports the RMSE and coefficient of determination calculated across all geometries.}
    \label{fig:parity}
\end{figure}

\Cref{fig:parity} compares surrogate predictions with EDGE2D--EIRENE values for five representative outputs: the carbon radiation front position, the \SI{3}{eV} electron temperature front position, the peak lower outer target electron temperature, the PID-controlled deuterium puff rate, and the total deuterium ion content.
The plotted points are drawn only from the held-out test set.

Overall, the parity plots demonstrate close agreement between the surrogate predictions and EDGE2D--EIRENE reference values, complementing the aggregate test set metrics in \cref{tab:performance}.
Nevertheless, the concentration of points about the one-to-one line is output- and value-dependent.
For the carbon front, the largest RMSE among the reference value bins occurs below $s_{\mathrm{norm}}\simeq0.58$, and errors decrease towards the X-point.
The peak target temperature and total content residuals are larger at high reference values, whereas puff rate errors are non-monotonic and largest in the lowest reference value quintile.
Near the target, the carbon front predictions have RMSE \num{0.0268} for $s_{\mathrm{norm}}\leq0.05$, but a mean positive bias of \num{0.0053} and $R^2=-8.02$ in this narrow-range subset.
The latter reflects its small intrinsic variance and indicates limited resolution of variations when the front is very close to the target.
The corresponding \SI{3}{eV} front subset has RMSE \num{0.0063} and $R^2=0.692$.
The unconstrained front models yield \num{90} carbon front and \num{27} \SI{3}{eV} front predictions below zero, all within these near-target subsets, with no predictions above one.
Simple post-processing can constrain such values.
The geometry colours reveal output-specific differences: CD has the largest RMSE for peak target temperature (\SI{0.180}{eV}, compared with \SI{0.026}{eV} for ED and \SI{0.016}{eV} for SXD) and for total content ($3.35\times10^{18}$, compared with $1.54\times10^{18}$ and $2.69\times10^{18}$, respectively).
However, the target temperature range also differs strongly between geometries (CD: \SIrange{0.64}{19.50}{eV}; ED: \SIrange{0.61}{2.61}{eV}; SXD: \SIrange{0.58}{1.43}{eV}), so these raw RMSE differences do not isolate an intrinsic geometry effect.

The vertical banding in parity plots of values measured along a flux tube, such as carbon radiation front position and \SI{3}{eV} electron temperature front position, arises from the discrete sampling of the computational grid.

\begin{figure}[htbp]
    \centering
    \includegraphics[width=0.92\linewidth]{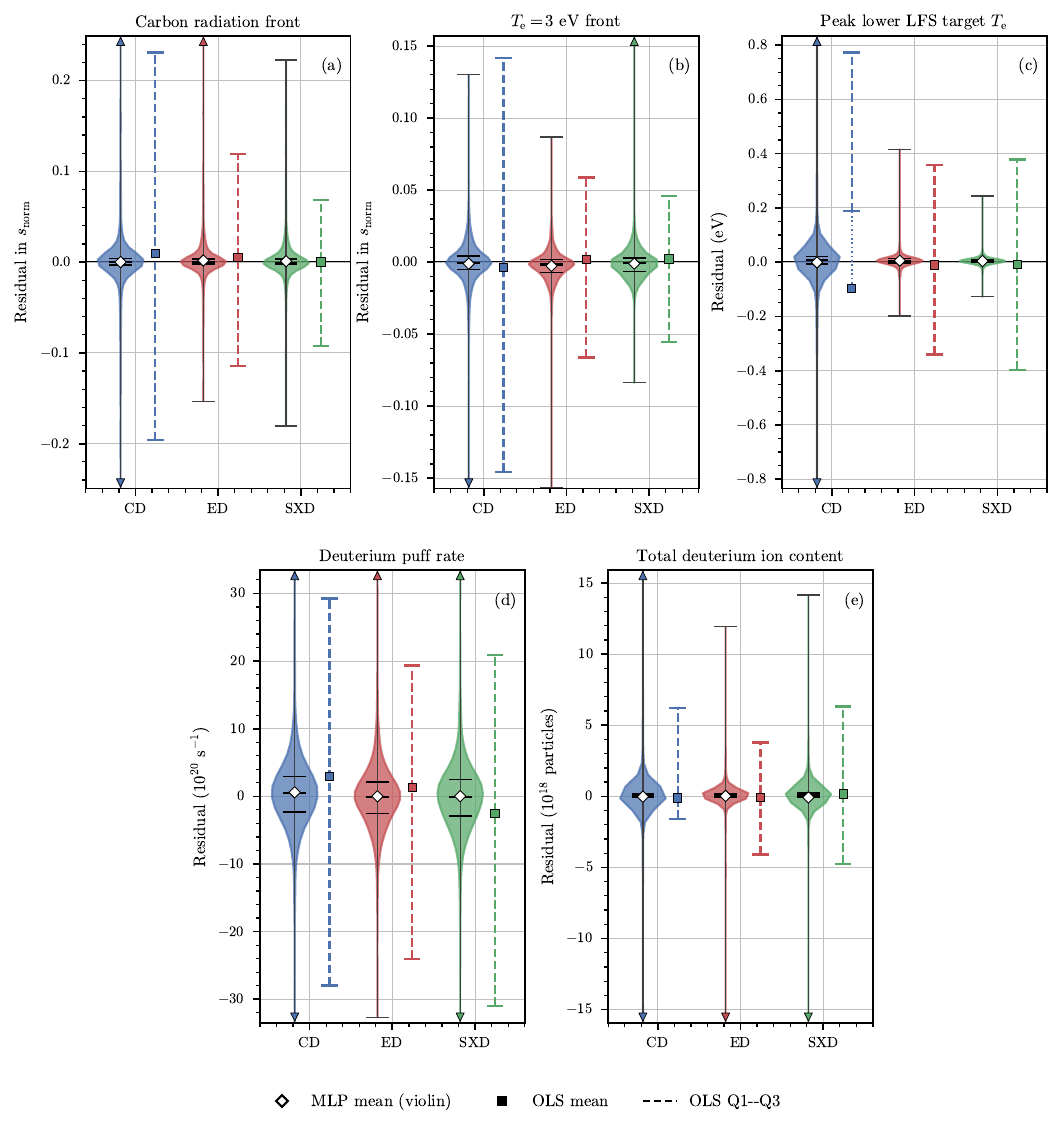}
    \caption{Signed residual distributions for the same held-out samples ($n=4676$) and representative outputs as \cref{fig:parity}, separated by conventional (CD), elongated (ED) and Super-X (SXD) divertor configuration.
    A residual is the surrogate prediction minus the EDGE2D--EIRENE value.
    For each configuration, the MLP distribution is shown on the left: the central black line denotes the median, the flanking black lines denote the first and third quartiles, and the white diamond denotes the mean.
    To its right, the OLS residual mean is shown as a coloured square, with horizontal caps at the first and third quartiles joined by a dashed line.
    A dotted extension connects a mean outside the interquartile interval to the nearest quartile, as occurs for CD in panel (c) because of its long negative residual tail.
    Each symmetric ordinate range contains at least the central 99\% of MLP residuals for its output and all OLS means and quartiles.
    Each coloured arrow tip indicates one or more MLP residuals beyond that range.}
    \label{fig:residuals}
\end{figure}

The signed residual distributions make the configuration-specific contributions to the aggregate metrics visible, as shown in \cref{fig:residuals} for the CD, ED and SXD subsets separately.
Here, a residual is defined as the surrogate prediction minus the EDGE2D--EIRENE value.
The OLS residual means and first and third quartiles are shown alongside the MLP distributions for comparison.
The CD OLS residual distributions for peak target temperature and total deuterium ion content have long negative tails, which pull their means below their medians.
For peak target temperature, the mean lies below the first quartile.
Positive and negative values denote overprediction and underprediction, respectively, while displacement of the distribution from zero indicates bias.
For the MLP, the central distributions of both front positions are close to zero in all three configurations.
The most marked difference is for peak target temperature.
Its central 95\% residual interval is \SIrange{-0.247}{0.199}{eV} for CD, compared with \SIrange{-0.039}{0.045}{eV} for ED and \SIrange{-0.021}{0.030}{eV} for SXD.
The larger variation in CD is consistent with the wider reference range of temperatures in that configuration.
The puff rate interquartile range widths are similar across configurations, spanning \num{4.67e20}--\num{5.40e20}\,\si{s^{-1}}, while the total content distributions have modest configuration-specific median shifts ($+\num{5.23e16}$, $+\num{7.77e16}$ and $+\num{6.54e16}$ particles for CD, ED and SXD, respectively).
These are descriptive differences in the sampled test distributions and are not evidence of an intrinsic geometry effect, because the input and output distributions differ between configurations.

\section{Applications of the surrogate}
\label{sec:applications}

\subsection{Rapid scans of detachment onset and sensitivity for varying divertor geometry and fuelling}
\label{sec:scan}

A primary benefit of the surrogate is that the discrete EDGE2D--EIRENE database can be converted into a rapidly evaluable continuous interpolation over the sampled continuous inputs.
This capability is demonstrated using the carbon radiation front position.
The scan varies $\nsep$ while fixing the core boundary input power at \SI{1.5}{MW} and setting the nitrogen puff rate to zero.
Separate curves are evaluated for each deuterium fuelling location in the CD, ED and SXD configurations.
Each curve is evaluated at \num{256} equally spaced densities from \SI{5e18}{m^{-3}} to \SI{1.5e19}{m^{-3}}.

\begin{figure}[htbp]
    \centering
    \includegraphics[width=0.74\linewidth]{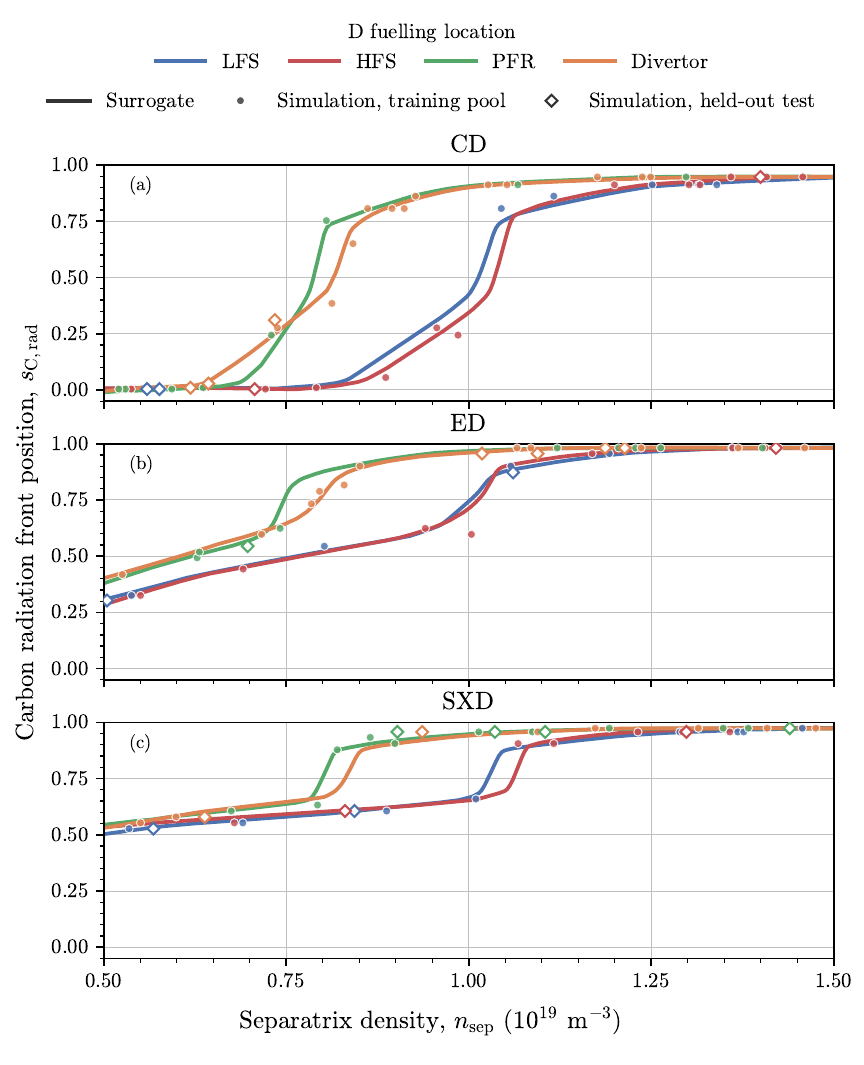}
    \caption{Surrogate predictions of normalised carbon radiation front position as a function of $\nsep$ for (a) the conventional divertor (CD), (b) the elongated divertor (ED) and (c) the Super-X divertor (SXD) at $\Pin=\SI{1.5}{MW}$ and $\GN=0$.
    Curves distinguish LFS, HFS, PFR and divertor deuterium fuelling, with \num{256} equally spaced density values per curve over \SIrange{5e18}{1.5e19}{m^{-3}}.
    Markers show converged, unseeded EDGE2D--EIRENE cases ($n=128$) with $\SI{1.475}{MW}\leq\Pin\leq\SI{1.525}{MW}$, coloured by fuelling location.
    Filled circles denote cases in the outer training pool (optimisation and validation subsets) and open diamonds denote held-out test cases ($n=25$).
    The vertical coordinate runs from the target at zero to the X-point at unity.}
    \label{fig:carbon-front}
\end{figure}

\Cref{fig:carbon-front} shows the resulting surrogate scans.
The plotted quantity is the carbon radiation front position defined in section~\ref{sec:surrogate-outputs}, expressed as normalised poloidal distance from the target to the X-point.
Across all twelve curves, increasing density produces an expected net displacement of the predicted front towards the X-point.
At the lowest density, the predictions are close to the divertor target in CD ($s_{\mathrm{C,rad}} \in \left[-0.011,0.007\right]$), further from the target in ED ($s_\mathrm{C,rad} \in \left[0.28, 0.40\right]$) and further still in SXD ($s_\mathrm{C,rad} \in \left[0.50, 0.54\right]$).
The small negative CD predictions reflect the unconstrained regression described in \cref{sec:representative-accuracy}.
The ED front is farther towards the X-point than the corresponding CD front at every sampled density and fuelling location.
The same ordering holds for SXD except at three HFS-fuelled scan points near $\nsep=\SI{1.06e19}{m^{-3}}$, where SXD lies below CD by at most 0.021, smaller than the global carbon front RMSE.
ED and SXD are not uniformly ordered over the scan.
The PFR and divertor fuelling curves move upstream at lower densities than the LFS and HFS curves through the transition in all three configurations, which is consistent with behaviour reported in~\textcite{lee2026} and~\textcite{xia2026a}.
The maximum spread between fuelling locations decreases from $0.75$ in CD to $0.36$ in ED and $0.30$ in SXD.

To compare the curves with the underlying simulations, \cref{fig:carbon-front} also shows every converged, unseeded EDGE2D--EIRENE case in the database with $\nsep$ inside the scan interval and $|\Pin-\SI{1.5}{MW}|\leq\SI{0.025}{MW}$.
Each case is compared with the curve for its configuration and fuelling location, evaluated at the same $\nsep$.
Because $\Pin$ is sampled continuously, no database case lies exactly on a scan trajectory, so a finite power tolerance is required.
The selected tolerance is 1.7\% of the scan power.
The predicted front displacement due to each case's power offset has median magnitude 0.0008 and maximum 0.039.
Of the overlaid cases, \num{80} were used for gradient-based optimisation, \num{23} belong to the validation subset used for learning rate reduction and early stopping, and \num{25} belong to the held-out test set.
Only the held-out cases provide an independent check, and they are distinguished in the figure.

The front transition is defined here as the range $0.1\leq s_{\mathrm{C,rad}}\leq0.9$ of the surrogate curve, over which the front position changes rapidly with density.
Outside this range, the curves agree closely with the overlaid simulations: the RMSE between curve and simulation is 0.010 for all 79 such cases and 0.015 for the 17 held-out cases among them.
Within the transition, the corresponding RMSEs are 0.042 for 49 cases and 0.025 for 8 held-out cases.
All 10 cases that differ from the curve by more than 0.05 lie in this range, with a maximum absolute difference of 0.125.
Across the 25 held-out cases, the curve-to-simulation RMSE is 0.019.
Evaluating the surrogate at each case's own $\Pin$ reduces it to 0.017, below the global held-out carbon front RMSE of 0.0303 in \cref{tab:performance}.
The overlaid simulations are therefore consistent with the scan trends, while the density interval over which the front moves rapidly towards the X-point is the least well predicted part of each trajectory.

The gradual upstream displacement of the carbon radiation maximum with increasing $\nsep$ in the CD scan agrees with the expected motion of the outer divertor total radiation peak towards the X-point during density ramps~\cite{federici2025}.

At the lowest scanned density, the carbon maximum remains near the target in CD but lies upstream in ED and SXD.
This contrast resembles the experimental comparison by Verhaegh et al., in which ED and SXD ionisation fronts remained separated from the target throughout the density ramps, while the CD front left the target only as core density increased~\cite{verhaegh2025}.

In the SXD scan, the carbon front reaches $s_{\mathrm{C,rad}}=0.75$ at $\nsep=\SI{8.31e18}{m^{-3}}$ under divertor fuelling, compared with $\SI{1.03e19}{m^{-3}}$ under LFS fuelling.
The lower density required under divertor fuelling agrees qualitatively with the SOLPS-ITER comparison by Xia et al., whose SXD simulations placed the ionisation fronts inferred from synthetic Fulcher emission at similar locations for divertor and LFS fuelling, with a lower upstream separatrix density in the divertor fuelled case~\cite{xia2026}.

\subsection{Nitrogen seeding scans}
\label{sec:nitrogen-scan}

The same approach is applied to nitrogen seeding, for which each additional seeding rate would otherwise require a separate EDGE2D--EIRENE calculation, with convergence becoming more expensive as the seeding rate increases (section~\ref{sec:sampling}).
The scan fixes $\nsep=\SI{7.5e18}{m^{-3}}$, $\Pin=\SI{1.5}{MW}$ and LFS deuterium fuelling for all equilibria.
This density lies below the carbon front transition of the unseeded CD scan in \cref{fig:carbon-front}, so that the response to seeding can be followed from attached CD conditions.
For each configuration and nitrogen puff location, $\GN$ is evaluated at 256 logarithmically spaced values from \SI{1.1e19}{s^{-1}} to \SI{1.95e21}{s^{-1}}, a range covered by the training data for every combination of configuration and puff location.
An unseeded prediction ($\GN=0$) is evaluated at the same density, power and fuelling location as a reference.
The emulated outputs are the carbon radiation and \SI{3}{eV} front positions, the peak lower LFS target electron temperature and the core boundary electron temperature.

\begin{figure}[htbp]
    \centering
    \includegraphics[width=\linewidth]{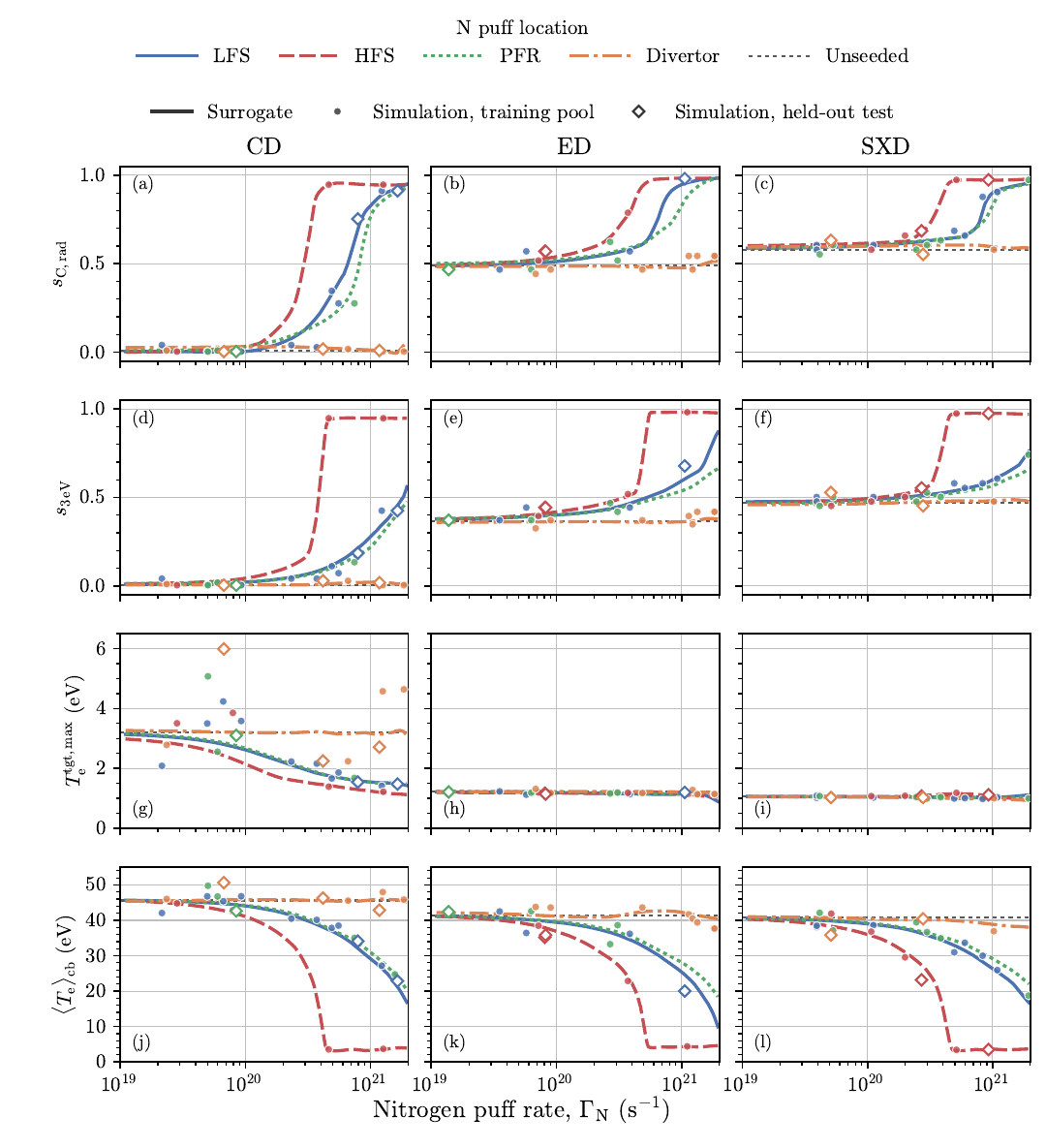}
    \caption{Surrogate predictions of response to the nitrogen puff rate $\GN$ at $\nsep=\SI{7.5e18}{m^{-3}}$, $\Pin=\SI{1.5}{MW}$ and LFS deuterium fuelling.
    Rows show (a)--(c) the carbon radiation front position, (d)--(f) the $\Te=\SI{3}{eV}$ front position, (g)--(i) the peak lower LFS target electron temperature and (j)--(l) the core boundary electron temperature.
    Columns correspond to the CD, ED and SXD divertor configurations.
    Colours and line styles distinguish the LFS, HFS, PFR and divertor nitrogen puff locations, and each curve is evaluated at 256 logarithmically spaced rates.
    The horizontal grey line shows the unseeded prediction.
    Markers show nearby converged EDGE2D--EIRENE cases ($n=70$) with LFS deuterium fuelling, $\SI{6.5e18}{m^{-3}}\leq\nsep\leq\SI{8.5e18}{m^{-3}}$ and $\SI{1.3}{MW}\leq\Pin\leq\SI{1.7}{MW}$, coloured by nitrogen puff location.
    Filled circles denote the outer training pool (optimisation and validation subsets); open diamonds denote held-out test cases ($n=14$).
    Front positions run from the target at zero to the X-point at unity.}
    \label{fig:nitrogen-scan}
\end{figure}

\Cref{fig:nitrogen-scan} shows that the seeding rate required for a given front displacement depends more strongly on the nitrogen puff location than on the divertor configuration.
HFS seeding produces the most abrupt change in all three magnetic equilibrium configurations.
Both fronts move close to the X-point, with the carbon radiation front reaching $s_{\mathrm{C,rad}}=0.9$ at $\GN = \bigl(\numrange{3.8}{4.5}\bigr)\times 10^{20}\,\si{\per\second}$.
Over the interval $\SI{2.5e20}{s^{-1}}\leq\GN\leq\SI{5.9e20}{s^{-1}}$, the separatrix electron temperature falls from \SIrange{28}{32}{eV} to \SIrange{3.2}{4.0}{eV}.
LFS and PFR seeding produce more gradual displacement: $s_{\mathrm{C,rad}}=0.9$ is reached only for $\GN\geq\SI{8.0e20}{s^{-1}}$, and $\langle\Te\rangle_{\mathrm{cb}}$ remains between \SIlist{9.9;22.1}{eV} at the highest rate.
The \SI{3}{eV} front does not reach $s_{3\,\mathrm{eV}}=0.9$ for either location in any configuration.
In CD it remains below 0.57.
Divertor seeding has little effect over the complete scan.
The CD fronts remain close to the target ($s_{\mathrm{C,rad}}<0.06$ and $s_{3\,\mathrm{eV}}<0.022$), the ED and SXD carbon radiation fronts move by less than 0.03, and $\langle\Te\rangle_{\mathrm{cb}}$ decreases by at most \SI{2.7}{eV} relative to the unseeded value.

The configurations differ in their unseeded starting points.
Without seeding, the carbon radiation front is at the target in CD ($s_{\mathrm{C,rad}}=0.009$) but at $s_{\mathrm{C,rad}}=0.49$ in ED and 0.58 in SXD.
The corresponding peak target electron temperature is \SI{3.2}{eV} in CD and \SIlist{1.2;1.1}{eV} in ED and SXD, respectively.
HFS, LFS and PFR seeding reduce the CD target temperature to \SIrange{1.1}{1.5}{eV}, whereas in ED and SXD it remains between \SIlist{0.9;1.3}{eV} over the entire scan.
Thus, in CD nitrogen seeding substantially reduces the target temperature, whereas in ED and SXD its main effects are to displace the fronts towards the X-point and, at sufficiently high rates, to cool the core boundary.

The same ordering, with HFS seeding acting at the lowest rates and divertor seeding having the weakest effect, persists when the scan is repeated for the other three deuterium fuelling locations and at $\nsep=\SI{1e19}{m^{-3}}$ with LFS fuelling.
For all 15 combinations of configuration and scan condition, including the primary scan, the HFS-seeded $\langle\Te\rangle_{\mathrm{cb}}$ falls to half its unseeded value at $\GN=\SIrange{2.6e20}{4.2e20}{s^{-1}}$, whereas divertor seeding never produces this reduction.
The converged simulations support these trends.
Among the seeded database cases with $\SI{6.5e18}{m^{-3}}\leq\nsep\leq\SI{8.5e18}{m^{-3}}$ and $\SI{1.3}{MW}\leq\Pin\leq\SI{1.7}{MW}$, irrespective of deuterium fuelling location, all 31 HFS-seeded cases with $\GN\geq\SI{5e20}{s^{-1}}$ have $\langle\Te\rangle_{\mathrm{cb}}<\SI{10}{eV}$.
None of the corresponding cases with other puff locations has $\langle\Te\rangle_{\mathrm{cb}}<\SI{10}{eV}$.
For the 65 held-out cases in this window, the surrogate RMSEs at their own inputs are 0.042 in $s_{\mathrm{C,rad}}$, 0.0095 in $s_{3\,\mathrm{eV}}$, \SI{0.033}{eV} in $\Te^{\mathrm{tgt,max}}$ and \SI{0.31}{eV} in $\langle\Te\rangle_{\mathrm{cb}}$.
The local front errors are of the same order as the global test metrics, while the two local temperature errors are smaller than their global values in \cref{tab:performance}.

\subsection{Computational cost}
\label{sec:cost}

The final inference implementation was benchmarked on an AMD Ryzen Threadripper 9960X CPU and an NVIDIA RTX PRO 5000 Blackwell GPU.
The timed region includes physical input preprocessing and host--device transfers but excludes model loading and input construction.
Single-point timings (with a batch size of one) use 20 warm-up evaluations and 200 measured evaluations.
Dense scan timings use one warm-up and five measured evaluations with batches of \num{65536} operating points per model.
Results are reported in \cref{tab:cost}.

Single-point latency is \SI{1.864}{\milli\second} for one model and \SI{20.63}{\milli\second} for all twelve models on the CPU.
The corresponding GPU times are \SI{1.819}{\milli\second} and \SI{21.74}{\milli\second}.
GPU execution therefore provides little latency benefit at batch size one.
For the dense benchmark, separatrix density and input power are each sampled on a 256-point linear grid over the ranges in \cref{tab:inputs}, repeated for all three geometries and four deuterium fuelling locations under unseeded conditions.
The resulting \num{786432} operating points require \SI{3.175}{\second} for one scalar model and \SI{38.92}{\second} for all twelve models on the CPU, compared with \SI{0.1422}{\second} and \SI{1.763}{\second}, respectively, on the GPU.

For comparison, a separate 35-point scan used fresh-start EDGE2D--EIRENE calculations on four cores of the same CPU, converging with a median wall time of \SI{2.61}{\hour} (IQR \SIrange{1.74}{3.48}{\hour}).

\begin{table}[htbp]
\centering
\caption{Measured wall time for direct simulation and surrogate inference, reported as median (IQR).
The CPU is an AMD Ryzen Threadripper 9960X and the GPU is an NVIDIA RTX PRO 5000 Blackwell.
The EDGE2D--EIRENE row contains 35 converged fresh-start calculations and uses four CPU cores per calculation.
Surrogate CPU rows use one thread.
Dense scan batches contain \num{65536} operating points per model.}
\label{tab:cost}
\small
\begin{tabular}{@{}p{0.31\linewidth}p{0.11\linewidth}p{0.24\linewidth}p{0.25\linewidth}@{}}
\toprule
Calculation & Batch & Hardware & Median wall time (IQR) \\
\midrule
EDGE2D--EIRENE fresh start & -- & CPU, 4 cores & \SI{2.61}{\hour} (\SIrange{1.74}{3.48}{\hour}) \\
One scalar, one point & 1 & CPU, 1 thread & \SI{1.864}{\milli\second} (1.849--\SI{1.888}{\milli\second}) \\
One scalar, one point & 1 & GPU & \SI{1.819}{\milli\second} (1.811--\SI{1.831}{\milli\second}) \\
All 12 scalars, one point & 1 & CPU, 1 thread & \SI{20.63}{\milli\second} (20.48--\SI{22.31}{\milli\second}) \\
All 12 scalars, one point & 1 & GPU & \SI{21.74}{\milli\second} (21.56--\SI{21.87}{\milli\second}) \\
One scalar, dense scan & \num{65536} & CPU, 1 thread & \SI{3.175}{\second} (3.155--\SI{3.188}{\second}) \\
One scalar, dense scan & \num{65536} & GPU & \SI{0.1422}{\second} (0.1421--\SI{0.1426}{\second}) \\
All 12 scalars, dense scan & \num{65536} & CPU, 1 thread & \SI{38.92}{\second} (38.88--\SI{38.97}{\second}) \\
All 12 scalars, dense scan & \num{65536} & GPU & \SI{1.763}{\second} (1.758--\SI{1.765}{\second}) \\
\bottomrule
\end{tabular}
\end{table}

\section{Discussion}
\label{sec:discussion}

\subsection{Capabilities demonstrated by the surrogate}
\label{sec:demonstrated}

The results demonstrate that a simple MLP can emulate a heterogeneous set of scalar EDGE2D--EIRENE outputs over the sampled MAST-U operating domain.
No architecture search or target-specific hyperparameter optimisation is required, and the same network topology and training procedure are used for all twelve observables.
This observation is consistent with recent SOLPS surrogate studies, which have likewise found fully connected neural networks to be a strong baseline for edge plasma emulation~\cite{dasbach2023}.
The linear regression baseline shows that this accuracy does not follow from a simple input--output relationship.
With the same inputs and preprocessing, OLS regression gives $R^2<0.9$ for ten of the twelve outputs and test RMSEs larger than those of the MLP by factors of 3.0--12.2.
The comparison therefore shows that a nonlinear model is needed to reach this accuracy and that the fixed MLP configuration is sufficient to attain $R^2\geq0.9462$ for every reported output.

The rapid carbon front and nitrogen seeding scans illustrate how a surrogate changes the practical use of such a database.
Direct EDGE2D--EIRENE calculations provide isolated samples, whereas the trained model permits dense evaluation of geometry, fuelling, seeding and operating point dependence at effectively negligible computational cost.

Relative to earlier surrogate work, the contribution of the present study is not the general observation that neural networks can emulate SOL simulations.
One-dimensional UEDGE and Hermes-3 surrogates have already demonstrated rapid detachment-oriented prediction~\cite{zhu2022,holt2024}, while recent two-dimensional UEDGE work has generated large detachment databases and control-oriented surrogates~\cite{zhu2025,zhao2026}.
Likewise, SOLPS studies have considered both reduced-fidelity interpolation and higher-dimensional neural surrogates~\cite{dasbach2023}.
The present database instead combines fully coupled EDGE2D--EIRENE calculations with kinetic neutral transport, three experimentally realisable MAST-U divertor geometries, localised deuterium fuelling and nitrogen seeding, and self-consistent carbon treatment.

\subsection{Scope and limitations}
\label{sec:limitations}

Several limitations define the range over which the reported test accuracy should be interpreted.
First, the held-out test errors measure emulation of EDGE2D--EIRENE, not prediction error against experiment.
Nevertheless, the published comparisons discussed in \cref{sec:model} show that two-dimensional edge simulations can reproduce selected measurements.
The surrogate therefore inherits both the physical assumptions and any discrepancies of those simulations.

Second, the random 80/20 outer split evaluates interpolation within the sampled distribution.
It does not test extrapolation to unseen divertor geometries, values outside the ranges in \cref{tab:inputs}, or different physical assumptions.
This distinction is particularly important for geometry: CD, ED and SXD are supplied to the MLP as categorical labels, so the network is not given a representation from which it could infer continuous geometry.
Future work could replace the categorical label with engineered geometric descriptors or spatial representations capable of supporting interpolation between equilibria.

Third, the anomalous cross-field transport coefficients are fixed across the database.
This reduces the dimensionality of the present problem and isolates the effects of the chosen geometry and operational inputs, but it excludes an important source of uncertainty in edge plasma modelling~\cite{dekeyser2022,fu2026}.
Extending the database to include transport variation is therefore a high priority for future surrogate and uncertainty quantification studies.

Finally, the present models predict scalar post-processed quantities.
They do not reconstruct target profiles, two-dimensional plasma fields, neutral distributions, or similar quantities.
Scalar models provide a low-complexity test of emulation across the complete database and are already useful for many parameter studies.
The archived simulations nevertheless contain substantially richer information, and the post-processing and training framework permits additional scalar surrogates to be generated readily.
Extension to profile and two-dimensional field prediction is left to future work.

\subsection{Future applications}
\label{sec:future}

The combination of a large simulation database and inexpensive surrogate inference enables several applications beyond those demonstrated here.
Dense parameter studies could be extended from the single-input scans shown in this work to multi-input trajectories, while global sensitivity analysis and uncertainty propagation could quantify the relative influence of geometry, density, input power and source location.
Surrogates could also support optimisation and automated scenario design, where direct coupling to EDGE2D--EIRENE would otherwise require prohibitive numbers of simulations.
With application-specific assessment against real-world measurements, millisecond-scale inference may also be useful for real-time prediction or model predictive control studies~\cite{lore2023}.

Future work should expand the input representation as well as the physical domain.
Continuous descriptors of magnetic geometry could allow interpolation between equilibria, while varying cross-field transport would expose the surrogate to a major modelling uncertainty.
Cross-field transport as an input would also enable rapid optimisation of the transport model needed to achieve a given experimental profile, which is required for interpretative simulations.
Profile and two-dimensional field surrogates would support applications requiring spatial information and could be explored using dimensionality reduction methods, operator learning approaches or other structured output architectures.
Active or adaptive sampling could further reduce the number of costly EDGE2D--EIRENE evaluations required to improve coverage in regions where the present database is sparse.

The public release of the database is intended to facilitate such work independently of the MLP architecture reported here.
In addition to supporting new physical studies, the same converged EDGE2D--EIRENE cases can serve as a common benchmark for comparing surrogate techniques and data-efficient training strategies.

\section{Conclusions}
\label{sec:conclusions}

A database of \num{38412} EDGE2D--EIRENE calculations has been generated across representative MAST-U conventional, elongated and Super-X divertor geometries, fuelling configurations, input powers, separatrix densities, and nitrogen seeding rates and puff locations.
Automated convergence screening identifies \num{23379} converged cases for surrogate training and evaluation.
Independent multi-layer perceptrons are used to emulate twelve representative scalar quantities spanning divertor front positions, target conditions, core boundary state, fuelling requirement and global deuterium ion content.

On a held-out random test set, the surrogate models achieve $0.9462\leq R^2\leq0.9994$ across the twelve reported outputs, with root-mean-square error values of 0.0303 in normalised carbon radiation front position and \SI{0.108}{eV} in peak lower low-field side target electron temperature.
An ordinary least-squares linear regression baseline using the same inputs and preprocessing attains only $0.0766\leq R^2\leq0.9879$, with test root-mean-square errors 3.0--12.2 times larger, showing that these outputs cannot be represented accurately as linear functions of the inputs.

The results demonstrate interpolation of heterogeneous EDGE2D--EIRENE outputs within the sampled operating domain without output-specific architecture optimisation.
In a fixed power, unseeded scan across conventional, elongated and Super-X divertor configurations, dense evaluation of the carbon radiation front surrogate predicts upstream motion with increasing density, generally farther-upstream fronts in elongated and Super-X divertor configurations than the conventional configuration, and earlier motion under private flux region or divertor fuelling than under low-field or high-field side fuelling, illustrating the use of the model for rapid parameter studies.
Converged simulations near the scan conditions agree with these curves with an RMSE of 0.010 away from the front transition and 0.042 within it.
Across the 25 held-out cases, the curve-to-simulation RMSE is 0.019.
At fixed density and power, nitrogen seeding scans predict that high-field side seeding moves both fronts close to the X-point and abruptly reduces the core boundary electron temperature over rates of approximately \SIrange{3e20}{6e20}{s^{-1}} in all three configurations, whereas divertor seeding has little effect.
Nearby converged simulations show the same distinction.
Warm single-point inference for one scalar requires \SI{1.864}{\milli\second} on one CPU thread and \SI{1.819}{\milli\second} on GPU, while the GPU evaluates all twelve outputs over the \num{786432}-point dense benchmark in \SI{1.763}{\second}.
By comparison, converged fresh-start EDGE2D--EIRENE calculations have a median wall time of \SI{2.61}{\hour} on four CPU cores.

The present surrogates should be interpreted as emulators of EDGE2D--EIRENE under the fixed transport assumptions represented in the database.
Future work will extend the database to additional transport assumptions and operating regimes, develop representations of continuous magnetic geometry, and move from scalar observables towards profile and two-dimensional field prediction.
More generally, these results show that surrogate modelling can make high-fidelity edge simulations accessible to workflows for which repeated direct EDGE2D--EIRENE evaluation is presently too expensive.

\section*{Acknowledgements}

We would like to thank Adriano Agnello of the STFC Hartree Centre for his helpful input during this work.
We would also like to thank David Moulton of the UK Atomic Energy Authority for providing information on the MAST-U operating space and discussions of the manuscript.

This work has been part-funded by the EPSRC Fusion Grant 2022/27 [grant number EP/W006839/1]. To obtain further information on the data and models underlying this paper please contact PublicationsManager@ukaea.uk.

This work was part-funded by the Fusion Computing Lab collaboration between UKAEA and STFC Hartree Centre.

\appendix
\section{Automated convergence classification}
\label{app:convergence}

The convergence classifier operates independently on the flux surface averaged ion temperature at the core boundary, the flux surface averaged electron temperature at the core boundary, and the total deuterium ion content in the EDGE2D computational domain.
One binary logistic regression model is trained for each quantity.
Training labels are derived from a small independent representative dataset for which convergence times were annotated manually.
Signal windows ending before the annotated convergence time are labelled pre-convergence, and windows at or after that time are labelled post-convergence.

For a simulation under assessment, only samples after the saved restart time are considered.
A nominal grace period of 150 samples is required before classification.
Each signal is smoothed using a 50-sample moving average, and the final 100 smoothed samples are retained.
For a signal $x$, the scalar convergence feature is
\begin{equation}
    g_x = \operatorname{mean}\left(\left|\frac{\nabla x}{\operatorname{mean}(x)}\right|\right),
\end{equation}
and the corresponding logistic regression input is
\begin{equation}
    z_x = \log_{10}\left(g_x + 10^{-10}\right).
\end{equation}
A signal is classified as converged when the predicted post-convergence probability satisfies
\begin{equation}
    P(\mathrm{post\text{-}convergence}\mid z_x) \geq 0.4.
\end{equation}
A simulation is retained only if all three signals satisfy this condition.
The classifier is used as a pragmatic automated filter for the large simulation campaign.

\clearpage

\printbibliography

\end{document}